\documentclass[aps,prl,amsmath,amssymb,reprint]{revtex4-1}

\usepackage{graphicx}
\usepackage{bm}
\usepackage{amsmath,empheq,amsfonts,amssymb} 
\usepackage{mathtools} 
\usepackage{mathrsfs} 
\usepackage[defaultlines=4,all]{nowidow}

\usepackage{lmodern}
\usepackage{color}
\usepackage{bm}
\usepackage{soul}
\usepackage{float}
\usepackage{tabularx}
\usepackage[T1]{fontenc} 
\usepackage{xcolor}
\usepackage{fancyref}
\usepackage{placeins} 
\usepackage{flafter}  
\usepackage{soul}
\usepackage{color}
\makeatletter
\let\Hy@backout\@gobble
\makeatother
\usepackage[version=3]{mhchem}
\usepackage{xcolor}
\usepackage{titlesec}
\titlespacing*{\section}{0pt}{1.1\baselineskip}{\baselineskip}

\begin{document}

	\author{D. Davidovikj$^1$}
	\email{d.davidovikj@tudelft.nl}
	\author{D.Bouwmeester$^1$}
	\author{H. S. J. van der Zant$^1$}
	\author{P. G. Steeneken$^{1,2}$}
	
	\affiliation{$^1$Kavli Institute of Nanoscience, Delft University of Technology, Lorentzweg 1, 2628 CJ Delft, The Netherlands \\
		$^2$Department of Precision and Microsystems Engineering, Delft University of Technology, Mekelweg 2, 2628 CD, Delft, The Netherlands} 
	\title{Graphene gas pumps}

\begin{abstract}
We report on the development of a pneumatically coupled graphene membrane system, comprising of two circular cavities connected by a narrow trench. Both cavities and the trench are covered by a thin few-layer graphene membrane to form a sealed dumbbell shaped chamber. Local electrodes at the bottom of each cavity allow for actuation  of each membrane separately, enabling electrical control and manipulation of the gas flow inside the channel. Using laser interferometry, we measure the displacement of each drum at atmospheric pressure, as a function of the frequency of the electrostatic driving force and provide a proof-of-principle of using graphene membranes to pump attolitre quantities of gases at the nanoscale.
\end{abstract}
\maketitle

Pumps have been of importance for humanity since early civilization. The Egyptians used a contraption called "shadoof" to take out water from the Nile that was used for irrigation. As technology progressed, better pumps usually meant higher pressure, larger flow, and hence, higher power. Micro- and nanofluidics in the past thirty years have substantially changed the way these devices are benchmarked. Microscale pumps are an essential ingredient in a microfluidic system, and the rapid advancements of biosciences require continually more devices capable of accurate micromixing and microdosing. This, in turn,  imposes better controllability, better accuracy, lower operational power, and much smaller flow rates~\cite{laser04,iverson08,lee10}. With respect to the first electrostatically actuated membrane pumps \cite{judy91,zengerle92}, that were presented more than 25 years ago, a tremendous reduction in size has been achieved. Pumps are also of interest for driving pneumatic actuators in micro- and nanoelectromechanical motors. The properties of graphene, like its atomic scale thickness and extreme flexibility, are very promising for further miniaturization of such nanofluidic devices.

\begin{figure}[h]
	
	\includegraphics[width=8.5cm]{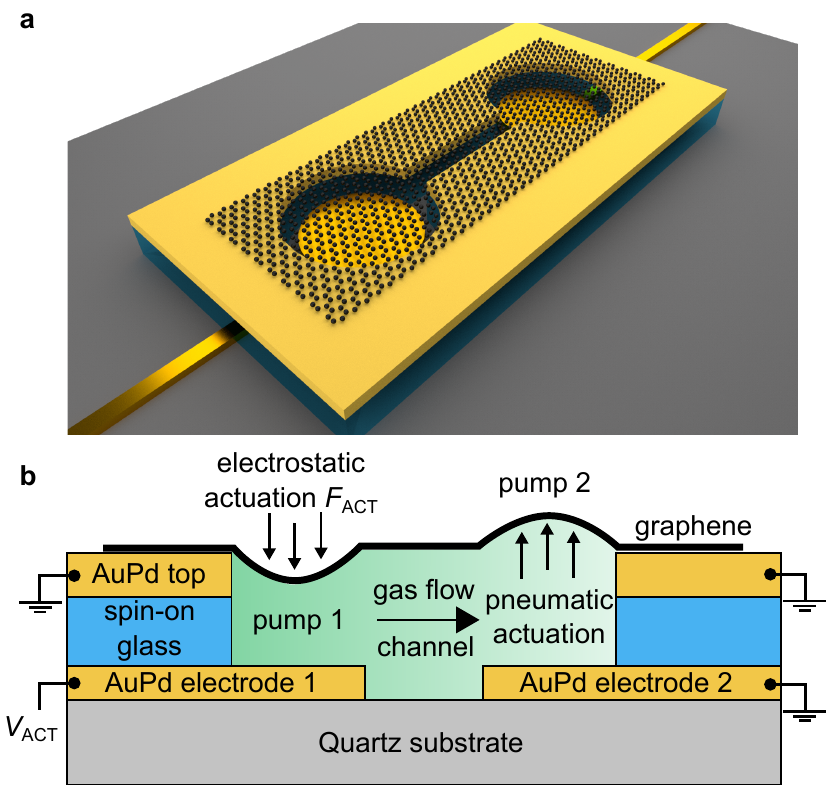}
	\caption{\label{fig:pu-Fig1} \textbf{Working principle of the pump.} \textbf{a,} A 3D schematic of the device: the graphene flake is covering two circular cavities that are connected through a narrow trench. \textbf{b,} Schematic cross-section of the pumps and actuation mechanism for the case that pump 1 is actuated.}
\end{figure}

Since the first realization of mechanical graphene devices~\cite{bunch07}, suspended 2D materials have attracted increasing attention in the MEMS/NEMS communities. Many device concepts have been proposed, including pressure sensors \cite{smith13pressure,dolleman15}, gas sensors \cite{bunch12,dolleman16}, mass sensors \cite{sakhaee08mass,atalaya10}, and graphene microphones~\cite{todorovic15,zhou15}. The high tension and low mass of graphene membranes have also inspired their implementation as high-speed actuators in micro-loudspeakers~\cite{zhou13}. Another attractive aspect of graphene membranes is their hermeticity~\cite{bunch08} and the ability to controllably introduce pores that are selectively permeable to gases~\cite{bunch12}. Although gas damping forces limit graphene's Q-factor at high frequencies, they provide a useful but little explored route towards graphene pumps and nanofluidics. For efficient pumps and pneumatics it is essential that most of the available power is used to move and pressurize the fluid, while minimizing the power required to accelerate and flex the pump membrane while minimizing leakage of the fluid outside of the system. In these respects, the low mass and high flexibility, combined with the impermeability of graphene membranes \cite{bunch08} provide clear advantages.

In this work we realise a system of two nanochambers (with a total volume of 7 fl) coupled by a narrow trench and sealed using a few-layer graphene flake. By designing a chip with individually accessible electrodes we construct a graphene micropump, capable of manipulating the gas flow between the two chambers using small driving voltages ($V_\mathrm{dc} \le 1$ V). Increasing the gas pressure in one of the nanochambers results in pneumatic actuation of the graphene drum that covers the other nanochamber via the connecting gas channel. To measure the displacement of the drums, we use laser interferometry and demonstrate successful pumping of gas between the two pneumatically coupled graphene nanodrums.

\begin{figure}[H]
	\centering
	\includegraphics[width=8.5cm]{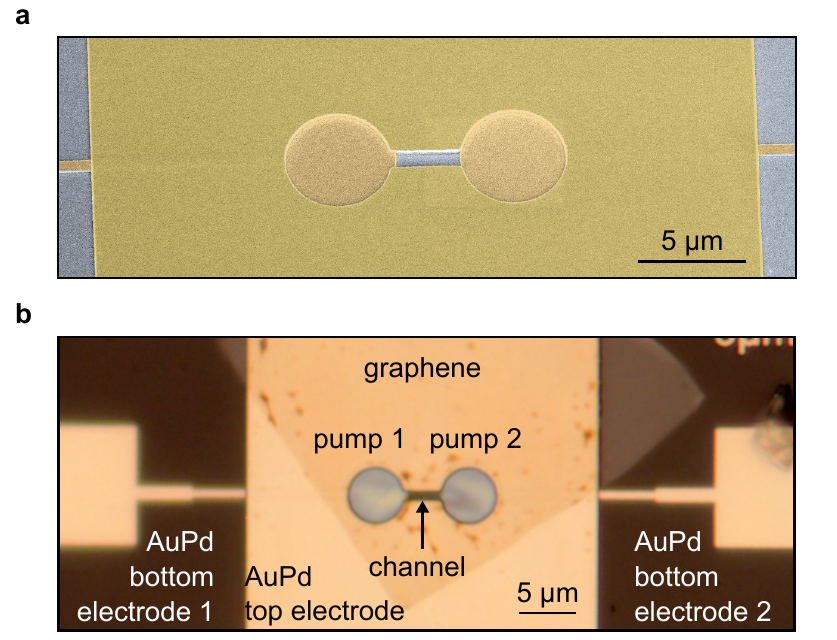}
	\caption{\label{fig:pu-Fig2} \textbf{Images of the fabricated device.} \textbf{a,} Scanning electron microscopy (SEM) image of the device prior to graphene transfer. \textbf{b,} Optical micrograph of the device after graphene transfer. }
\end{figure}

\section{Device description}
The device concept is presented in Fig.~\ref{fig:pu-Fig1}a. Two circular AuPd electrodes (thickness: 60 nm) at the cavity bottom (one for addressing each of the membranes) are separated by a thin layer (130 nm) of spin-on-glass (SOG) silicon oxide from the metallic (AuPd) top electrode (thickness: 85 nm). The few-layer (FL) graphene flake (black), with a thickness of 4 nm, is in direct electrical contact with this top electrode. The entire device is fabricated on top of a quartz substrate to minimize capacitive cross-talk. The device fabrication is described in detail in~\cite{davidovikj17}. A cross-section along the direction of the trench of the device is shown in Fig.~\ref{fig:pu-Fig1}b, which illustrates the working principle. The actuation voltage $V_\mathrm{ACT,1}$ is applied between AuPd electrode 1 and ground, while keeping AuPd electrode 2 and the AuPd top electrode grounded. As a result, pump 1 experiences an electrostatic force $F_\mathrm{ACT}$, causing it to deflect downward. This compresses the gas underneath the membrane and the induced pressure difference causes a gas flow through the channel between the two nanochambers. This results in a pressure increase that causes the other membrane (pump 2) to bulge upward.

Figure~\ref{fig:pu-Fig2}a shows a false-coloured SEM image of the device after fabrication. The AuPd is shown in light (bottom electrodes) and dark (top electrode) yellow. The diameter of each drum is 5 $\mu$m and the trench connecting them is 1 $\mu$m wide and 3 $\mu$m long. Figure~\ref{fig:pu-Fig2}b shows an optical image of the measured device. The image shows the two bottom electrodes, together with the top metallic island on which the dumbbell shape is patterned. Graphene is transferred last (as described in~\cite{davidovikj17}) and it is visible in the image as a darker area on top of the metallic island.

\begin{figure}[H]
	\centering
	\includegraphics[width=8.5cm]{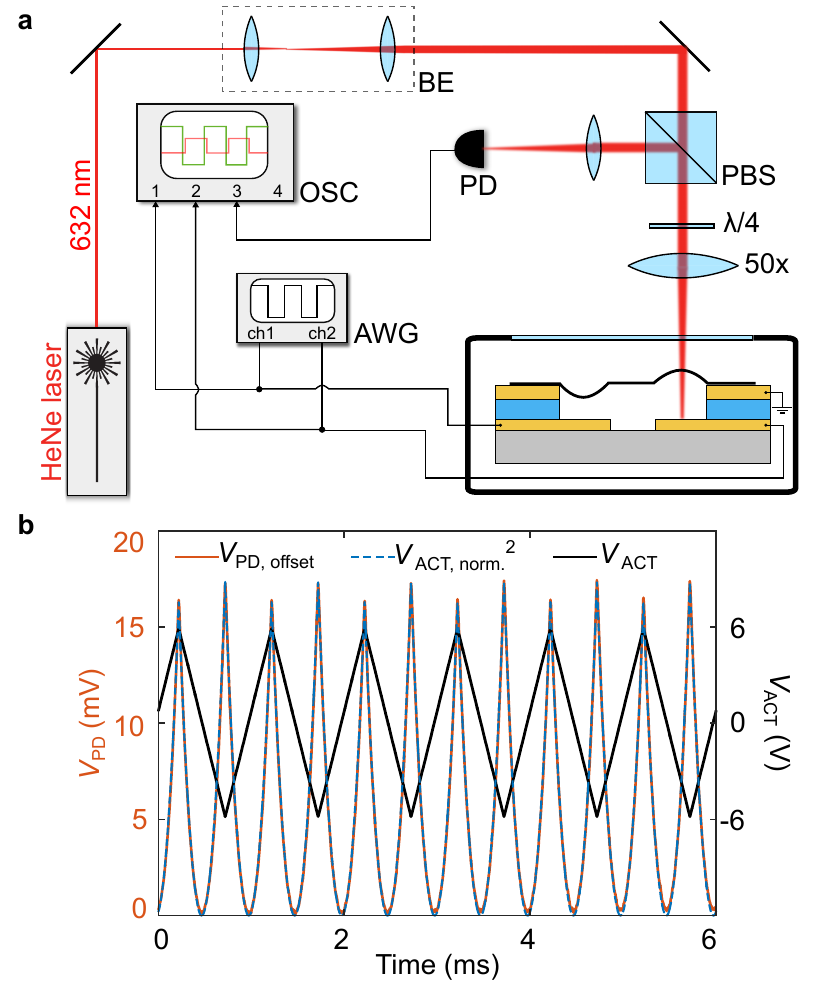}
	\caption{\label{fig:pu-Fig3} \textbf{Measurement setup.} \textbf{a,} Setup for electrostatic actuation and interferometric motion readout of either of the pumps. PD: photodiode, OSC: oscilloscope, AWG: arbitrary waveform generator, BE: beam expander and PBS: polarized beam splitter. \textbf{b,} Offset photodiode voltage (red curve) for a triangular input signal $V_\mathrm{ACT}$ (black curve). The dashed blue curve represents the input voltage squared, normalized to the photodiode voltage: $V_\mathrm{ACT,norm.}^2 = \alpha (V_\mathrm{ACT} + \beta)^2 $. The term $\beta=-\,0.13$ V accounts for residual charge on the graphene flake~\cite{chen09}.}
\end{figure}

\section{Readout}
The readout of the drum motion of the is performed using a laser interferometer, shown schematically in Fig.~\ref{fig:pu-Fig3}a. A red HeNe laser is focused on one of the graphene membranes, and the sample is mounted in a pressure chamber in a \ce{N2} environment at ambient pressure and room temperature. When the membrane moves, the optical interference between the light reflected from the bottom electrode and the light reflected from the graphene causes the light intensity on the photodiode detector to depend strongly on the drum position. By lateral movement of the laser spot, the motion of either of the pumps can be detected. The photodiode signal is read out via an internal first-order low-pass filter with a cut-off frequency of 50 kHz.

\begin{figure*}[t]
	\centering
	\includegraphics[width=13cm]{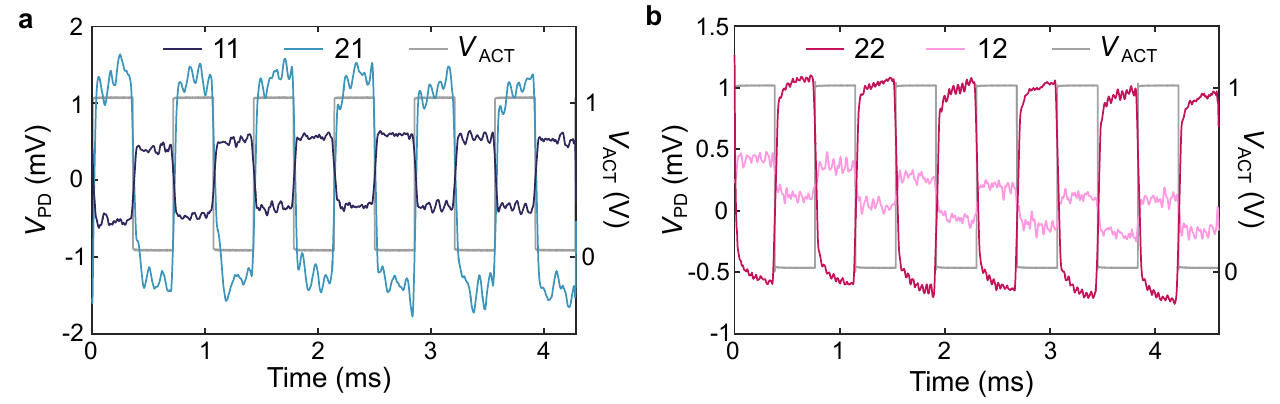}
	\caption{\label{fig:pu-Fig4} \textbf{Time domain measurements.} \textbf{a,} Measured displacement (photodiode voltage) of pump 1 when actuating pump 1 (measurement 11) and drum 2 (measurement 21). \textbf{b,} Measured displacement (photodiode voltage) of drum 2 when actuating pump 1 (measurement 12) and drum 2 (measurement 22). The actuation voltage is shown on the right y-axis. Measurements of each of the drums are performed at constant laser position to ensure that the transduction of the system ($V_\mathrm{PD}/z$) is constant.}
\end{figure*}

For electrostatic actuation, the two bottom electrodes are connected to two channels of an arbitrary waveform generator, where one is grounded and the other one is actuated (Figs. \ref{fig:pu-Fig1}b and \ref{fig:pu-Fig3}). The actuation voltage ($V_\mathrm{ACT}$) on each of the drums and the photodiode voltage ($V_\mathrm{PD}$) are measured using an oscilloscope. The top electrode (i.e., the graphene flake) is electrically grounded during the measurements. Since there are 2 pumps that can be actuated (pump 1 and pump 2) and either of them can be detected with the red laser there are 4 measurement configurations indicated by $V_\mathrm{PD,11}$,$V_\mathrm{PD,21}$,$V_\mathrm{PD,12}$ and $V_\mathrm{PD,22}$, where the first number indicates the pump that is actuated and the second number indicates the pump that is read out. 

We first characterize the responsivity of the system by applying a triangular voltage signal to one of the drums while measuring its motion with the laser. The measurement is shown in Fig.~\ref{fig:pu-Fig3}b. The force acting on the drum scales quadratically with $V_\mathrm{ACT}$ and therefore, for small amplitudes, it is expected that the amplitude of the drum would also depend quadratically on $V_\mathrm{ACT}$ (assuming $F_\mathrm{ACT}  = -kz$, see Supporting Information Section II). The fact that the voltage read out from the photodiode matches the scaled square of the input voltage (blue curve in Fig.~\ref{fig:pu-Fig3}b) confirms that the assumption of linear transduction ($V_\mathrm{ACT}^2\propto V_\mathrm{PD}$) of the motion is valid.

\section{Gas pump and pneumatic actuation}

Pneumatic actuation is one of the most efficient ways to transfer force over large distances in small volumes. At the microscale, the pneumatic coupling also has the advantage of converting the attractive downward electrostatic force on pump 1 to an upward force on the graphene membrane of pump 2 (Fig.~\ref{fig:pu-Fig1}b). Thus, proof for gas pumping and pneumatic actuation can be obtained by detecting that the drums move in opposite directions.  

The drums are actuated using a square-wave voltage input $V_\mathrm{AC,p-p} = 1$ V with a frequency of 1.3 kHz, plotted in Fig.~\ref{fig:pu-Fig4}a and Fig.~\ref{fig:pu-Fig4}b (grey curves). Figure~\ref{fig:pu-Fig4}a shows a measurement of the displacement of pump 1, when applying $V_\mathrm{ACT}$ on pump 1 while keeping pump 2 grounded (dark blue curve) or when actuating pump 2 while keeping pump 1 grounded (light blue curve). Both curves show a main frequency component that is coinciding with the frequency of the driving signal, meaning that the detected motion is a consequence of the applied actuation. However, when switching the actuation to pump 2 it is seen that the photodiode voltage ($V_\mathrm{PD,21}$) is 180 degrees out of phase with respect to $V_\mathrm{PD,11}$. This is indicative of an out-of-phase motion of the two drums. Such effect is possible only if the actuation of pump~1 (in the 21 configuration) is pneumatic, i.e., mediated by gas displacement from one chamber to the other. 

The same experiments are repeated in Fig.~\ref{fig:pu-Fig4}b when moving the laser spot to pump~2. The red curve represents the case when pump~2 is electrically actuated while keeping pump~1 grounded and the pink curve represents the case when pump~1 is electrically actuated and pump~2 is kept grounded. The same conclusion can be drawn: the two curves are 180 degrees out-of-phase, confirming that the drums move in opposite directions.

The differences in signal amplitudes in Fig.~\ref{fig:pu-Fig4} are attributed to differences in the effective cavity depths between the pumps that affect the actuation/detection efficiency (this may happen due to morphological imperfections in the graphene flake). To confirm that the coupling is mediated by gas, the experiment is repeated at low pressure. After keeping the sample at 0.1 mbar for 48 hours, the gas is completely evacuated from the cavity~\cite{bunch08}. In this case no sign of motion of the second drum is observed in the $V_\mathrm{PD,12}$ signal, showing that pneumatic actuation is absent in vacuum (see Supporting Information Section I).

\begin{figure*}[t]
	\includegraphics[width=13cm]{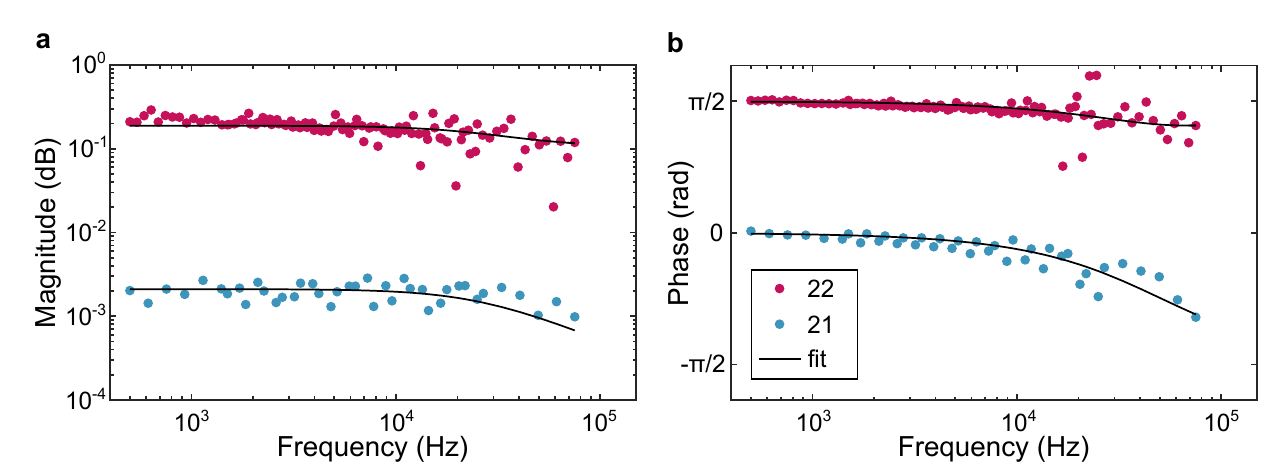}
	\caption{\label{fig:pu-Fig5} \textbf{Frequency domain measurements.} Bode plots (\textbf{a,} magnitude and \textbf{b,} phase) of the system for actuation of pump 2. The data points are coloured according to the measurement scheme: the red points represent actuation and detection at pump 2, while the light blue represent actuation at pump 2 and measurement of pump 1. The fit is according to the model represented with Equations~\eqref{pu-eq1} and~\eqref{pu-eq2}.}
\end{figure*}


Assuming that the cavities are hermetically sealed by the graphene (valid for very low permeation rates~\cite{bunch08}), the pneumatically coupled graphene pump system can be modelled in the quasi-static regime by a set of two linear differential equations describing the pressure increase $\Delta P_i$ in each of the chambers. The pressure difference can then be related to the displacement $z_i$ of the drums (details of the model and the derivation are given in the Supporting Information Section I). In the frequency domain the solutions of these differential equations can be written in terms of the Fourier transforms $\mathscr{F}$ of the solutions: $z_1(\omega) =\mathscr{F}(z_1) $, $z_2(\omega) =\mathscr{F}(z_2) $ and $F(\omega) = c\mathscr{F}(V_2^2)$, where $F$ is the actuation pressure and $c$ is a function of the squeeze number and the gap size $g_0 = $ 155 nm. When the actuation signal is applied to pump 2, the response is given by:

\begin{equation}
\label{pu-eq1}
z_1(\omega) = \frac{1}{2}\frac{1}{1+i\omega \tau}\frac{A}{k}F(\omega);
\end{equation}

\begin{equation}
\label{pu-eq2}
z_2(\omega) = -\Big(\frac{1}{2}\frac{1}{1+i\omega \tau}+\frac{1}{a}\Big)\frac{A}{k}F(\omega),
\end{equation}

\noindent where $z_1$ and $z_2$ are the displacements of pump 1 and pump 2 respectively, $\omega$ is the actuation frequency, $A$ is the area of each drum, $k$ is the spring constant of the drums and $a$ is the squeeze number. The time constant $\tau$ is then given by:
\begin{equation}
\label{tau}
\tau = \frac{1 + a}{2b},
\end{equation}

\noindent where the constant $b$ is related to the gas flow through the channel. Assuming a laminar Poiseuille flow, $b$ is dependent on the geometry of the channel and the effective viscosity of the gas, in this case nitrogen (see Supporting Information Section II).

To investigate the nanoscale gas dynamics experimentally, the frequency response of the system is measured. The actuation voltage is applied on pump 2. The frequency of the square-wave input signal ($V_\mathrm{ACT} (t)$, see Fig.~\ref{fig:pu-Fig4}) is varied from 510 Hz to 23 kHz. For each actuation frequency, the Fourier transform is taken of both the input and output signal. By taking the ratio of the input and output at each of the driving frequencies a frequency response plot is obtained. We make use of the fact that the input square-wave contains higher harmonics to increase the amount of data acquired by a single time response signal, thereby increasing the frequency resolution.

The resulting Bode plots are shown in Fig.~\ref{fig:pu-Fig5}. It can be seen that both the magnitude and phase of the resulting frequency response curves are flat up to a frequency of 10 kHz. At higher frequencies the amplitude of the motion of the second drum drops, which suggests that at these frequencies the pumping efficiency starts to become limited by gas dynamics through the narrow channel. Fits using the model described by Equations~\eqref{pu-eq1} and~\eqref{pu-eq2} show that the response of the pumps correspond to a first-order RC low-pass filter with a characteristic time constant of $\tau= 39.3 \pm 3.4\,\mu$s, resulting in a cut-off frequency of 25.4~$\pm$~2.2 kHz. 

The demonstrated graphene-based pump system is not only of extraordinarily small size (total volume of 7 fl), but it is also capable of pumping very small amounts of gas: assuming the spring constant to be in the order of $k\approx$ 1 N/m, less than 80 al of gas is pumped through the channel each cycle. The thermal noise, due to charge fluctuations on the capacitor plates, sets a lower limit to the pump rate of less than 1 zl/$\sqrt{\mathrm{Hz}}$, which is equivalent to less than 30~\ce{N2}~molecules/$\sqrt{\mathrm{Hz}}$ at ambient pressure and room temperature. The maximum electrostatic pressure that can be generated by the graphene pump with the given geometry is 0.5 bar, limited by the breakdown voltage of the dielectric ($V_\mathrm{b}=$ 16 V). The typical force exerted at $V_\mathrm{ACT}=$ 1 V is 4 nN, corresponding to an electrostatic pressure of 2 mbar.

Besides the pneumatic actuation and pumping, the system also allows the study of gas dynamics in channels of sub-micron dimensions, where the free path length of molecules is smaller than the channel height, even at atmospheric pressure. By controllably introducing pores in the graphene, the graphene pump can be used for molecular sieving of gases, or even aspiration and dispensing of liquids. The presented system can therefore be used as a platform for studying anomalous viscous effects in narrow constrictions as well as graphene-gas interactions at the nanoscale. It thus provides a route towards scaling down nanofluidic systems by using graphene membranes coupled by nanometre-sized channels.

\section*{Acknowledgements}

This work was supported by the Netherlands Organisation for Scientific Research (NWO/OCW), as part of the Frontiers of Nanoscience (NanoFront) program and the European Union Seventh Framework Programme under grant agreement $\mathrm{n{^\circ}~696656}$ Graphene Flagship. Parts of this manuscript have been published in the form of a proceeding at the IEEE 31th International Conference on Micro Electro Mechanical Systems~\cite{davidovikj18pumps}.

\pagebreak
\onecolumngrid
\setcounter{figure}{0}
\renewcommand{\figurename}{FIG. S\!\!}

\newpage

\section*{Supporting Information: Graphene gas pumps}
\label{pu-appendix}
\subsection*{I. Model of the pump system}
\label{pu-model}
\subsection*{Equations of motion}
In this section the model for the two drum system will be explained, starting with the assumptions that were made in order to arrive at the model. 

The drums are modelled as simple harmonic oscillators. A parallel plate capacitor model is taken to model the electrostatic force on the drum, which holds for small deflections of the membrane with respect to gap. The gas inside the circular cavities is modelled as an ideal gas and gas inertia is neglected. The interactions of the gas are considered to be isothermal. Poiseuille flow through the trench between the two drums is considered. 

The mechanics of the drums are described using Newton's second law of motion. The forces that act on the drums are the tension force of the drums (assuming equal spring constants $k$ and masses $m$), the pressure force acting on the drum and the electrostatic forces coming from the charge stored in the membrane-electrode capacitor. No damping is considered apart from damping due to the gas pressure. The electrostatic force is applied to the first drum (pump 1). We name deflection of drum $i$ with respect to the gap $z_i$, such that a positive value of $z_i$ corresponds to the drum bulging upward. We consider the outside air to be at ambient pressure $P$, while the pressure inside chamber $i$ is $ P_i$. The pressure difference across the drum is be called $\Delta P_i$. The gap size is denoted as $g_0$, $\varepsilon_0$ is the vacuum permittivity and $A = r^2\pi$ is the area of each of the drums. In terms of these quantities the equations of motion are:

\begin{empheq}[left=\empheqlbrace]{align}
& m\frac{d^2z_1}{dt^2}=-kz_1+\Delta P_1 A -\frac{V^2\varepsilon_0 A}{2(g_0+z_1)^2}  \label{eq:N2d1};\\
& m\frac{d^2z_2}{dt^2}=-kz_2+\Delta P_2 A  \label{eq:N2d2} \,.
\end{empheq}

\begin{figure}[H]
	\centering
	\includegraphics[width=8.5cm]{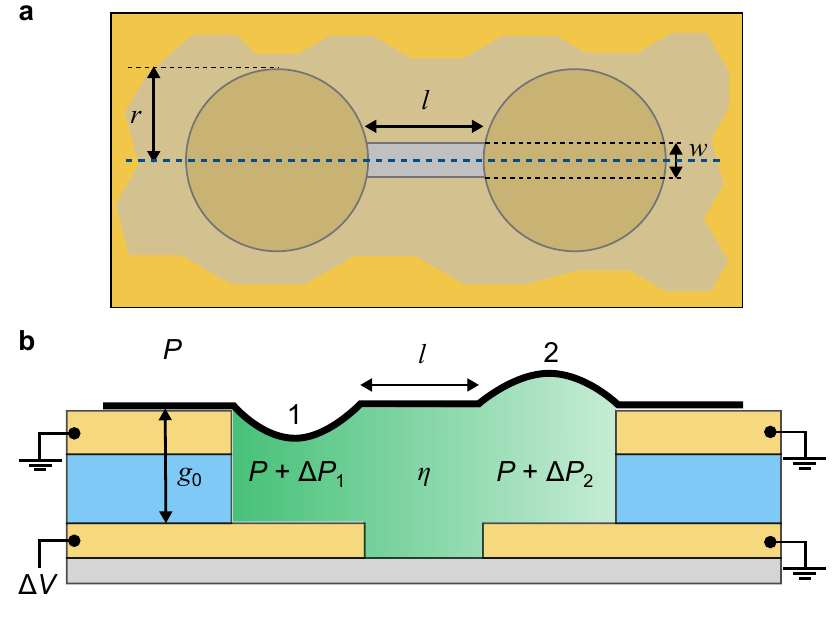}
	\caption{\label{fig:pu-FigS1} \textbf{Schematic of the device.} \textbf{a,} Top view of the pump system. \textbf{b,} A side view of the cross section through the dashed blue line in \textbf{a}. A potential $\Delta V$ is applied on pump 1 that results in the membrane compressing the gas, causing an increase of	pressure in the chamber $P_1 = P+\Delta P_2$. This causes gas flow through the channel and results in pump 2 bulging upward. Since the channel is smaller than the chamber, it will obstruct the flow of gas, hence the pressure in the other chamber $P_2 = P+\Delta P_2$ will lag.}
\end{figure}

With this, the mechanics are fully described. These equations have one driving force, the electrostatic force experienced by pump 1. The pressure force due to the gas is not a driving force and should react to the motion of the membrane. In order to describe the pressure in the drums, the ideal gas law is taken:
\begin{empheq}[left=\empheqlbrace]{align}
& P_1 = P+\Delta P_1=\frac{n_1\bar{R}T}{V_1}=\frac{n_1\bar{R}T}{A(g_0+z_1)};\\
& P_2 = P+\Delta P_2=\frac{n_2\bar{R}T}{V_2}=\frac{n_2\bar{R}T}{A(g_0+z_2)}\,.
\end{empheq}

The quantity $n_i$ in these equations stands for the amount of moles of gas in chamber~$i$.
These two equations, together with Equations \eqref{eq:N2d1} and \eqref{eq:N2d2} give a set of equations in which the membranes are coupled to the gas pressure in the drums. The pressures are now coupled to one another using the Poiseuille flow equation. This equation determines the rate of pressure induced flow of a viscous fluid across a channel. In the pump system, this fluid is the nitrogen gas in the cavity. The Poiseuille flow equation describes both $n_1$ and $n_2$ through the following differential equation:
\begin{equation}
\Phi=(1-0.63\frac{g_0}{w})\frac{(\Delta P_1-\Delta P_2) g_0^3 w}{12\eta l}=\frac{dn_2}{dt}\frac{V_2}{n_2} = \frac{dn_2}{dt}\frac{\bar{R}T}{P+\Delta P_2}=-\frac{dn_1}{dt}\frac{\bar{R}T}{P+\Delta P_1}\,.
\label{eq:pois}
\end{equation}

In this equation $\Phi$ is the volumetric flux of gas through the channel, $\bar{R}$ is the ideal gas constant, and $\eta$ is the dynamic viscosity of the gas.
The Poiseuille flow equation acts as the coupling between the two cavities. In using this equation to express the change in the amount of gas molecules in the cavities we have implicitly added the condition that the total amount of gas molecules in the pump system is conserved, which holds assuming no gas permeation outside the cavities.
In order to incorporate Equation~\eqref{eq:pois} into the model, the time derivatives of the ideal gas laws are taken:
\begin{empheq}[left=\empheqlbrace]{align}
& \frac{d\Delta P_1}{dt}V_1+(P+\Delta P_1)A\frac{dz_1}{dt}=\frac{dn_1}{dt}\bar{R}T;\\
& \frac{d\Delta P_2}{dt}V_2+(P+\Delta P_2)A\frac{dz_2}{dt}=\frac{dn_2}{dt}\bar{R}T\,.
\end{empheq}

Filling in Equation~\eqref{eq:pois} results in the following set of equations:

\begin{empheq}[left=\empheqlbrace]{align}
&\frac{d\Delta P_1}{dt}=-\frac{(P+\Delta P_1)}{g_0+z_1}\frac{dz_1}{dt} \notag\\
& +  (1-0.63\frac{g_0}{w})\frac{(\Delta P_2-\Delta P_1)(P+\Delta P_1) g_0^3 w}{12\eta lA(g_0+z_1)}\,; \\
& \frac{d\Delta P_2}{dt}=-\frac{(P+\Delta P_2)}{g_0+z_2}\frac{dz_2}{dt}\notag\\ 
&+ (1-0.63\frac{g_0}{w})\frac{(\Delta P_1-\Delta P_2)(P+\Delta P_2) g_0^3 w}{12\eta lA(g_0+z_2)} \,.
\end{empheq}

To neatly express the model of the two drum system, the four differential equations that describe the system are given together. The following set of equations describe the two drum system:
\begin{empheq}[left=\empheqlbrace]{align}
& \frac{d^2z_1}{dt^2}=-\frac{k}{m}z_1+ \frac{A}{m}\Delta P_1 -\frac{V^2\varepsilon_0 A}{2m(g_0+z_1)^2}\,; \label{eq:sys1}\\
& \frac{d^2z_2}{dt^2}=-\frac{k}{m}z_2+\frac{A}{m}\Delta P_2\,; \label{eq:sys2}\\
&\frac{d\Delta P_1}{dt}=-\frac{(P+\Delta P_1)}{g_0+z_1}\frac{dz_1}{dt} \notag\\
& +  (1-0.63\frac{g_0}{w})\frac{(\Delta P_2-\Delta P_1)(P+\Delta P_1) g_0^3 w}{12\eta lA(g_0+z_1)}\,; \\
& \frac{d\Delta P_2}{dt}=-\frac{(P+\Delta P_2)}{g_0+z_2}\frac{dz_2}{dt}\notag\\ 
&+ (1-0.63\frac{g_0}{w})\frac{(\Delta P_1-\Delta P_2)(P+\Delta P_2) g_0^3 w}{12\eta lA(g_0+z_2)} \,.
\end{empheq}

\subsection*{Quasi-static equations}
The quasi-static limit of these equations is taken.
In this case the second derivatives in Equations~\eqref{eq:N2d1} and~\eqref{eq:N2d2} are negligible. Newton's law is now equivalent to a force balance, indicating that at all times the drums are at an equilibrium position. This equilibrium position changes in time due to the changing gas pressure and voltage. As such there is still a response to the driving force. In order to find approximate solutions to the differential equations, Equations~\eqref{eq:sys1} and~\eqref{eq:sys2} are linearised. The force balance that is found is:

\begin{empheq}[left=\empheqlbrace]{align}
& kz_1= A\Delta P_1 -\frac{V^2\varepsilon_0 A}{2g_0^2}\,;\label{eq:qstat1}\\
& kz_2= A\Delta P_2\,.\label{eq:qstat2}
\end{empheq}

The linearised differential equations for the pressure are:
\begin{empheq}[left=\empheqlbrace]{align}
& \frac{d\Delta P_1}{dt}=-\frac{P}{g_0}\frac{dz_1}{dt}+(1-0.63\frac{g_0}{w})\frac{(\Delta P_2-\Delta P_1)P g_0^2 w}{12\eta lA}\,;\label{eq:preqstat3}\\
& \frac{d\Delta P_2}{dt}=-\frac{P}{g_0}\frac{dz_2}{dt}+(1-0.63\frac{g_0}{w})\frac{(\Delta P_1-\Delta P_2)P g_0^2 w}{12\eta lA}\,. \label{eq:preqstat4}
\end{empheq}

Now filling the force balance into this differential equation eliminates all displacement terms and yields the following differential equations for the pressure:
\begin{empheq}[left=\empheqlbrace]{align}
& \frac{d\Delta P_1}{dt}(1+\frac{PA}{kg_0})=(1-0.63\frac{g_0}{w})\frac{(\Delta P_2-\Delta P_1)P g_0^2 w}{12\eta lA}\notag\\
&+\frac{PA  }{kg_0}\frac{\varepsilon_0}{2g_0^2}\frac{dV^2}{dt}\,;\label{eq:qstat3}\\
& \frac{d\Delta P_2}{dt}(1+\frac{PA}{kg_0})=(1-0.63\frac{g_0}{w})\frac{(\Delta P_1-\Delta P_2)P g_0^2 w}{12\eta lA}\,. \label{eq:qstat4}
\end{empheq}

For simplicity, we define the following constants:
\begin{equation*}
a=\frac{PA}{kg_0}\,; 
\end{equation*}
\begin{equation*}
b=(1-0.63\frac{g_0}{w})\frac{P g_0^2 w}{12\eta lA}\,.
\end{equation*}

Here $a$ is the squeeze number and $b$ is related to the gas flow dynamics through the channel. This allows us to put the differential equations into the following simple form:

\begin{equation}
\frac{d}{dt}\begin{bmatrix}
\Delta P_1\\
\Delta P_2\\
\end{bmatrix}=\frac{-b}{1+a}\begin{bmatrix}
1 && -1\\
-1 && 1\\
\end{bmatrix}
\begin{bmatrix}
\Delta P_1\\
\Delta P_2\\
\end{bmatrix}+c\frac{d}{dt}\begin{bmatrix}
V^2\\
0\\
\end{bmatrix}\,;
\end{equation}

\begin{equation}
c=\frac{a}{1+a}\frac{\varepsilon_0}{2g_0^2}\,.
\end{equation}

\subsection*{Frequency spectrum of the system}
In order to investigate the behaviour of this differential equation, a Fourier transform of the differential equations is taken:
\begin{equation}
i\omega\begin{bmatrix}
\mathscr F (\Delta P_1)\\
\mathscr F (\Delta P_2)\\
\end{bmatrix}=\frac{-b}{1+a}\begin{bmatrix}
1 && -1\\
-1 && 1\\
\end{bmatrix}
\begin{bmatrix}
\mathscr F (\Delta P_1)\\
\mathscr F (\Delta P_2)\\
\end{bmatrix}+ci\omega\begin{bmatrix}
\mathscr F (V^2)\\
0\\
\end{bmatrix}.
\end{equation}

The frequency spectra found from these equations are given by

\begin{equation}
\mathscr F (\Delta P_1)=\frac{\frac{1}{2}+i\omega\tau}{1+i\omega\tau }c\mathscr F(V^2)\,;
\end{equation}
\begin{equation}
\mathscr F (\Delta P_2)=\frac{\frac{1}{2}}{1 +i\omega\tau}c \mathscr F(V^2)\,,
\end{equation}

\noindent with $\tau=\frac{1+a}{2b}$. This time also defines the cutoff frequency of the gas pump system $\omega_0=\frac{2b}{1+a}$. The function Fourier transform of $\Delta
P_2$ takes the form of a low pass filter. We are more interested in the Fourier transform of $P_1$, whose magnitude and phase of $\mathscr F (\Delta P_1)$ are given below.

\begin{equation}
|\mathscr F (\Delta P_1)|=\frac{1}{2}\sqrt{\frac{1+\omega^24\tau^2}{1+\omega^2\tau^2}}c|\mathscr F(V^2)|\,;
\end{equation}

\begin{equation}
\phi=Arg[\mathscr F (\Delta P_1)]=\arctan\Big(\frac{\omega\tau}{1+\omega^22\tau^2}\Big)\,.
\end{equation}
%
%

Finally, we examine the behaviour of the drum displacement. The algebraic equations found for the displacement allow us to directly calculate the Fourier transform of the displacement from the pressures and the square of the electrostatic potential.

\begin{equation}
\mathscr F(z_1)=-\Big(\frac{1}{2}\frac{1}{1+i\omega \tau}+\frac{1}{a}\Big)\frac{A}{k}c\mathscr F (V^2)\,;
\end{equation}

\begin{equation}
\mathscr F(z_2)=\frac{1}{2}\frac{1}{1+i\omega \tau}\frac{A}{k}c\mathscr F (V^2)\,.
\end{equation}

Once more it can be seen that $\mathscr F(z_2)$ is given by applying a low pass filter on the driving force. The magnitude and phase of $\mathscr F(z_1)$ and find:

\begin{equation}
|\mathscr F(z_1)|=\sqrt{\frac{1}{a^2}+\Big(\frac{1}{a}+\frac{1}{4}\Big)\frac{1}{1+\omega^2\tau^2}}\frac{A}{k}c|\mathscr F(V^2)|\,;
\end{equation}

\begin{equation}
\phi_1=Arg(\mathscr F(z_1))=\pi-\arctan\Big(\frac{a \omega\tau }{2+a+2\omega^2\tau^2}\Big)\,.
\label{eq:z1phase}
\end{equation}

We now use these equations to model the curves from Fig.~\ref{fig:pu-Fig5}.

%
%
\newpage
\subsection*{II. Measurement in vacuum}
\label{pu-vacuum}
\begin{figure}[H]
	\centering
	\includegraphics[width=6.5cm]{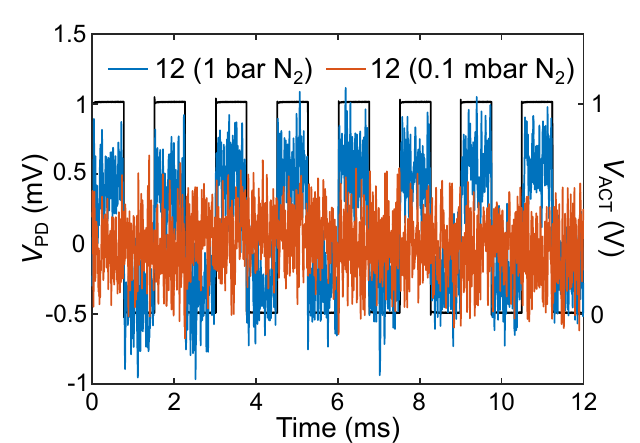}
	\caption{\label{fig:pu-vacuum} \textbf{Measurement in vacuum.} Measurement of the motion of pump 2 when actuating pump 1 at atmospheric pressure (blue curve) and in vacuum (orange curve).}
\end{figure}

A comparison of "12" measurements in vacuum and in \ce{N2} is shown in Fig.~\ref{fig:pu-vacuum}. In ambient pressure, the motion of pump 2 responds to the actuation of pump 1, mediated by the gas in the chamber. The absence of motion of pump 2 in vacuum (orange curve in Fig.~\ref{fig:pu-vacuum}) is another confirmation of pneumatic actuation in the system.

\newpage

%


\begin{thebibliography}{19}%
	\makeatletter
	\providecommand \@ifxundefined [1]{%
		\@ifx{#1\undefined}
	}%
	\providecommand \@ifnum [1]{%
		\ifnum #1\expandafter \@firstoftwo
		\else \expandafter \@secondoftwo
		\fi
	}%
	\providecommand \@ifx [1]{%
		\ifx #1\expandafter \@firstoftwo
		\else \expandafter \@secondoftwo
		\fi
	}%
	\providecommand \natexlab [1]{#1}%
	\providecommand \enquote  [1]{``#1''}%
	\providecommand \bibnamefont  [1]{#1}%
	\providecommand \bibfnamefont [1]{#1}%
	\providecommand \citenamefont [1]{#1}%
	\providecommand \href@noop [0]{\@secondoftwo}%
	\providecommand \href [0]{\begingroup \@sanitize@url \@href}%
	\providecommand \@href[1]{\@@startlink{#1}\@@href}%
	\providecommand \@@href[1]{\endgroup#1\@@endlink}%
	\providecommand \@sanitize@url [0]{\catcode `\\12\catcode `\$12\catcode
		`\&12\catcode `\#12\catcode `\^12\catcode `\_12\catcode `\%12\relax}%
	\providecommand \@@startlink[1]{}%
	\providecommand \@@endlink[0]{}%
	\providecommand \url  [0]{\begingroup\@sanitize@url \@url }%
	\providecommand \@url [1]{\endgroup\@href {#1}{\urlprefix }}%
	\providecommand \urlprefix  [0]{URL }%
	\providecommand \Eprint [0]{\href }%
	\providecommand \doibase [0]{http://dx.doi.org/}%
	\providecommand \selectlanguage [0]{\@gobble}%
	\providecommand \bibinfo  [0]{\@secondoftwo}%
	\providecommand \bibfield  [0]{\@secondoftwo}%
	\providecommand \translation [1]{[#1]}%
	\providecommand \BibitemOpen [0]{}%
	\providecommand \bibitemStop [0]{}%
	\providecommand \bibitemNoStop [0]{.\EOS\space}%
	\providecommand \EOS [0]{\spacefactor3000\relax}%
	\providecommand \BibitemShut  [1]{\csname bibitem#1\endcsname}%
	\let\auto@bib@innerbib\@empty
	\bibitem [{\citenamefont {Laser}\ and\ \citenamefont
		{Santiago}(2004)}]{laser04}%
	\BibitemOpen
	\bibfield  {author} {\bibinfo {author} {\bibfnamefont {D.~J.}\ \bibnamefont
			{Laser}}\ and\ \bibinfo {author} {\bibfnamefont {J.~G.}\ \bibnamefont
			{Santiago}},\ }\href {http://dx.doi.org/10.1088/0960-1317/14/6/R01}
	{\bibfield  {journal} {\bibinfo  {journal} {Journal of Micromechanics and
				Microengineering}\ }\textbf {\bibinfo {volume} {14}},\ \bibinfo {pages} {R35}
		(\bibinfo {year} {2004})}\BibitemShut {NoStop}%
	\bibitem [{\citenamefont {Iverson}\ and\ \citenamefont
		{Garimella}(2008)}]{iverson08}%
	\BibitemOpen
	\bibfield  {author} {\bibinfo {author} {\bibfnamefont {B.~D.}\ \bibnamefont
			{Iverson}}\ and\ \bibinfo {author} {\bibfnamefont {S.~V.}\ \bibnamefont
			{Garimella}},\ }\href {http://dx.doi.org/10.1007/s10404-008-0266-8}
	{\bibfield  {journal} {\bibinfo  {journal} {Microfluidics and Nanofluidics}\
		}\textbf {\bibinfo {volume} {5}},\ \bibinfo {pages} {145} (\bibinfo {year}
		{2008})}\BibitemShut {NoStop}%
	\bibitem [{\citenamefont {Lee}\ \emph {et~al.}(2010)\citenamefont {Lee},
		\citenamefont {An},\ and\ \citenamefont {Hunt}}]{lee10}%
	\BibitemOpen
	\bibfield  {author} {\bibinfo {author} {\bibfnamefont {S.}~\bibnamefont
			{Lee}}, \bibinfo {author} {\bibfnamefont {R.}~\bibnamefont {An}}, \ and\
		\bibinfo {author} {\bibfnamefont {A.~J.}\ \bibnamefont {Hunt}},\ }\href
	{http://dx.doi.org/10.1038/nnano.2010.81} {\bibfield  {journal} {\bibinfo
			{journal} {Nature Nanotechnology}\ }\textbf {\bibinfo {volume} {5}},\
		\bibinfo {pages} {412} (\bibinfo {year} {2010})}\BibitemShut {NoStop}%
	\bibitem [{\citenamefont {Judy}\ \emph {et~al.}(1991)\citenamefont {Judy},
		\citenamefont {Tamagawa},\ and\ \citenamefont {Polla}}]{judy91}%
	\BibitemOpen
	\bibfield  {author} {\bibinfo {author} {\bibfnamefont {W.}~\bibnamefont
			{Judy}}, \bibinfo {author} {\bibfnamefont {T.}~\bibnamefont {Tamagawa}}, \
		and\ \bibinfo {author} {\bibfnamefont {D.~L.}\ \bibnamefont {Polla}},\ }\href
	{http://dx.doi.org/10.1109/MEMSYS.1991.114792} {\bibfield  {journal}
		{\bibinfo  {journal} {Proceedings IEEE MEMS}\ ,\ \bibinfo {pages} {182}}
		(\bibinfo {year} {1991})}\BibitemShut {NoStop}%
	\bibitem [{\citenamefont {Zengerle}\ \emph {et~al.}(1992)\citenamefont
		{Zengerle}, \citenamefont {Richter},\ and\ \citenamefont
		{Sandmaier}}]{zengerle92}%
	\BibitemOpen
	\bibfield  {author} {\bibinfo {author} {\bibfnamefont {R.}~\bibnamefont
			{Zengerle}}, \bibinfo {author} {\bibfnamefont {A.}~\bibnamefont {Richter}}, \
		and\ \bibinfo {author} {\bibfnamefont {H.}~\bibnamefont {Sandmaier}},\ }\href
	{http://dx.doi.org/10.1109/MEMSYS.1992.187684} {\bibfield  {journal}
		{\bibinfo  {journal} {Proceedings IEEE MEMS}\ ,\ \bibinfo {pages} {19}}
		(\bibinfo {year} {1992})}\BibitemShut {NoStop}%
	\bibitem [{\citenamefont {Bunch}\ \emph {et~al.}(2007)\citenamefont {Bunch},
		\citenamefont {Van Der~Zande}, \citenamefont {Verbridge}, \citenamefont
		{Frank}, \citenamefont {Tanenbaum}, \citenamefont {Parpia}, \citenamefont
		{Craighead},\ and\ \citenamefont {McEuen}}]{bunch07}%
	\BibitemOpen
	\bibfield  {author} {\bibinfo {author} {\bibfnamefont {J.~S.}\ \bibnamefont
			{Bunch}}, \bibinfo {author} {\bibfnamefont {A.~M.}\ \bibnamefont {Van
				Der~Zande}}, \bibinfo {author} {\bibfnamefont {S.~S.}\ \bibnamefont
			{Verbridge}}, \bibinfo {author} {\bibfnamefont {I.~W.}\ \bibnamefont
			{Frank}}, \bibinfo {author} {\bibfnamefont {D.~M.}\ \bibnamefont
			{Tanenbaum}}, \bibinfo {author} {\bibfnamefont {J.~M.}\ \bibnamefont
			{Parpia}}, \bibinfo {author} {\bibfnamefont {H.~G.}\ \bibnamefont
			{Craighead}}, \ and\ \bibinfo {author} {\bibfnamefont {P.~L.}\ \bibnamefont
			{McEuen}},\ }\href {http://dx.doi.org/10.1126/science.1136836} {\bibfield
		{journal} {\bibinfo  {journal} {Science}\ }\textbf {\bibinfo {volume}
			{315}},\ \bibinfo {pages} {490} (\bibinfo {year} {2007})}\BibitemShut
	{NoStop}%
	\bibitem [{\citenamefont {Smith}\ \emph {et~al.}(2013)\citenamefont {Smith},
		\citenamefont {Vaziri}, \citenamefont {Niklaus}, \citenamefont {Fischer},
		\citenamefont {Sterner}, \citenamefont {Delin}, \citenamefont {{\"O}stling},\
		and\ \citenamefont {Lemme}}]{smith13pressure}%
	\BibitemOpen
	\bibfield  {author} {\bibinfo {author} {\bibfnamefont {A.}~\bibnamefont
			{Smith}}, \bibinfo {author} {\bibfnamefont {S.}~\bibnamefont {Vaziri}},
		\bibinfo {author} {\bibfnamefont {F.}~\bibnamefont {Niklaus}}, \bibinfo
		{author} {\bibfnamefont {A.}~\bibnamefont {Fischer}}, \bibinfo {author}
		{\bibfnamefont {M.}~\bibnamefont {Sterner}}, \bibinfo {author} {\bibfnamefont
			{A.}~\bibnamefont {Delin}}, \bibinfo {author} {\bibfnamefont
			{M.}~\bibnamefont {{\"O}stling}}, \ and\ \bibinfo {author} {\bibfnamefont
			{M.}~\bibnamefont {Lemme}},\ }\href
	{http://dx.doi.org/10.1016/j.sse.2013.04.019} {\bibfield  {journal} {\bibinfo
			{journal} {Solid-State Electronics}\ }\textbf {\bibinfo {volume} {88}},\
		\bibinfo {pages} {89 } (\bibinfo {year} {2013})}\BibitemShut {NoStop}%
	\bibitem [{\citenamefont {Dolleman}\ \emph
		{et~al.}(2016{\natexlab{a}})\citenamefont {Dolleman}, \citenamefont
		{Davidovikj}, \citenamefont {Cartamil-Bueno}, \citenamefont {van~der Zant},\
		and\ \citenamefont {Steeneken}}]{dolleman15}%
	\BibitemOpen
	\bibfield  {author} {\bibinfo {author} {\bibfnamefont {R.~J.}\ \bibnamefont
			{Dolleman}}, \bibinfo {author} {\bibfnamefont {D.}~\bibnamefont
			{Davidovikj}}, \bibinfo {author} {\bibfnamefont {S.~J.}\ \bibnamefont
			{Cartamil-Bueno}}, \bibinfo {author} {\bibfnamefont {H.~S.~J.}\ \bibnamefont
			{van~der Zant}}, \ and\ \bibinfo {author} {\bibfnamefont {P.~G.}\
			\bibnamefont {Steeneken}},\ }\href
	{http://dx.doi.org/10.1021/acs.nanolett.5b04251} {\bibfield  {journal}
		{\bibinfo  {journal} {Nano Letters}\ }\textbf {\bibinfo {volume} {16}},\
		\bibinfo {pages} {568} (\bibinfo {year} {2016}{\natexlab{a}})}\BibitemShut
	{NoStop}%
	\bibitem [{\citenamefont {Koenig}\ \emph {et~al.}(2012)\citenamefont {Koenig},
		\citenamefont {Wang}, \citenamefont {Pellegrino},\ and\ \citenamefont
		{Bunch}}]{bunch12}%
	\BibitemOpen
	\bibfield  {author} {\bibinfo {author} {\bibfnamefont {S.~P.}\ \bibnamefont
			{Koenig}}, \bibinfo {author} {\bibfnamefont {L.}~\bibnamefont {Wang}},
		\bibinfo {author} {\bibfnamefont {J.}~\bibnamefont {Pellegrino}}, \ and\
		\bibinfo {author} {\bibfnamefont {J.~S.}\ \bibnamefont {Bunch}},\ }\href
	{http://dx.doi.org/10.1038/nnano.2012.162} {\bibfield  {journal} {\bibinfo
			{journal} {Nature Nanotechnology}\ }\textbf {\bibinfo {volume} {7}},\
		\bibinfo {pages} {728} (\bibinfo {year} {2012})}\BibitemShut {NoStop}%
	\bibitem [{\citenamefont {Dolleman}\ \emph
		{et~al.}(2016{\natexlab{b}})\citenamefont {Dolleman}, \citenamefont
		{Cartamil-Bueno}, \citenamefont {van~der Zant},\ and\ \citenamefont
		{Steeneken}}]{dolleman16}%
	\BibitemOpen
	\bibfield  {author} {\bibinfo {author} {\bibfnamefont {R.~J.}\ \bibnamefont
			{Dolleman}}, \bibinfo {author} {\bibfnamefont {S.~J.}\ \bibnamefont
			{Cartamil-Bueno}}, \bibinfo {author} {\bibfnamefont {H.~S.}\ \bibnamefont
			{van~der Zant}}, \ and\ \bibinfo {author} {\bibfnamefont {P.~G.}\
			\bibnamefont {Steeneken}},\ }\href
	{http://dx.doi.org/10.1088/2053-1583/4/1/011002} {\bibfield  {journal}
		{\bibinfo  {journal} {2D Materials}\ }\textbf {\bibinfo {volume} {4}},\
		\bibinfo {pages} {011002} (\bibinfo {year} {2016}{\natexlab{b}})}\BibitemShut
	{NoStop}%
	\bibitem [{\citenamefont {Sakhaee-Pour}\ \emph {et~al.}(2008)\citenamefont
		{Sakhaee-Pour}, \citenamefont {Ahmadian},\ and\ \citenamefont
		{Vafai}}]{sakhaee08mass}%
	\BibitemOpen
	\bibfield  {author} {\bibinfo {author} {\bibfnamefont {A.}~\bibnamefont
			{Sakhaee-Pour}}, \bibinfo {author} {\bibfnamefont {M.}~\bibnamefont
			{Ahmadian}}, \ and\ \bibinfo {author} {\bibfnamefont {A.}~\bibnamefont
			{Vafai}},\ }\href {http://dx.doi.org/10.1016/j.ssc.2007.10.032} {\bibfield
		{journal} {\bibinfo  {journal} {Solid State Communications}\ }\textbf
		{\bibinfo {volume} {145}},\ \bibinfo {pages} {168} (\bibinfo {year}
		{2008})}\BibitemShut {NoStop}%
	\bibitem [{\citenamefont {Atalaya}\ \emph {et~al.}(2010)\citenamefont
		{Atalaya}, \citenamefont {Kinaret},\ and\ \citenamefont
		{Isacsson}}]{atalaya10}%
	\BibitemOpen
	\bibfield  {author} {\bibinfo {author} {\bibfnamefont {J.}~\bibnamefont
			{Atalaya}}, \bibinfo {author} {\bibfnamefont {J.~M.}\ \bibnamefont
			{Kinaret}}, \ and\ \bibinfo {author} {\bibfnamefont {A.}~\bibnamefont
			{Isacsson}},\ }\href {http://dx.doi.org/10.1209/0295-5075/91/48001}
	{\bibfield  {journal} {\bibinfo  {journal} {Europhysics Letters}\ }\textbf
		{\bibinfo {volume} {91}},\ \bibinfo {pages} {48001} (\bibinfo {year}
		{2010})}\BibitemShut {NoStop}%
	\bibitem [{\citenamefont {Todorovi\'{c}}\ \emph {et~al.}(2015)\citenamefont
		{Todorovi\'{c}}, \citenamefont {Matkovi\'{c}}, \citenamefont
		{Mili\'{c}evi\'{c}}, \citenamefont {Jovanovi\'{c}}, \citenamefont
		{Gaji\'{c}}, \citenamefont {Salom},\ and\ \citenamefont
		{Spasenovi\'{c}}}]{todorovic15}%
	\BibitemOpen
	\bibfield  {author} {\bibinfo {author} {\bibfnamefont {D.}~\bibnamefont
			{Todorovi\'{c}}}, \bibinfo {author} {\bibfnamefont {A.}~\bibnamefont
			{Matkovi\'{c}}}, \bibinfo {author} {\bibfnamefont {M.}~\bibnamefont
			{Mili\'{c}evi\'{c}}}, \bibinfo {author} {\bibfnamefont {D.}~\bibnamefont
			{Jovanovi\'{c}}}, \bibinfo {author} {\bibfnamefont {R.}~\bibnamefont
			{Gaji\'{c}}}, \bibinfo {author} {\bibfnamefont {I.}~\bibnamefont {Salom}}, \
		and\ \bibinfo {author} {\bibfnamefont {M.}~\bibnamefont {Spasenovi\'{c}}},\
	}\href {http://dx.doi.org/10.1088/2053-1583/2/4/045013} {\bibfield  {journal}
		{\bibinfo  {journal} {2D Materials}\ }\textbf {\bibinfo {volume} {2}},\
		\bibinfo {pages} {045013} (\bibinfo {year} {2015})}\BibitemShut {NoStop}%
	\bibitem [{\citenamefont {Zhou}\ \emph {et~al.}(2015)\citenamefont {Zhou},
		\citenamefont {Zheng}, \citenamefont {Onishi}, \citenamefont {Crommie},\ and\
		\citenamefont {Zettl}}]{zhou15}%
	\BibitemOpen
	\bibfield  {author} {\bibinfo {author} {\bibfnamefont {Q.}~\bibnamefont
			{Zhou}}, \bibinfo {author} {\bibfnamefont {J.}~\bibnamefont {Zheng}},
		\bibinfo {author} {\bibfnamefont {S.}~\bibnamefont {Onishi}}, \bibinfo
		{author} {\bibfnamefont {M.}~\bibnamefont {Crommie}}, \ and\ \bibinfo
		{author} {\bibfnamefont {A.~K.}\ \bibnamefont {Zettl}},\ }\href
	{http://dx.doi.org/10.1073/pnas.1505800112} {\bibfield  {journal} {\bibinfo
			{journal} {Proceedings of the National Academy of Sciences}\ }\textbf
		{\bibinfo {volume} {112}},\ \bibinfo {pages} {8942} (\bibinfo {year}
		{2015})}\BibitemShut {NoStop}%
	\bibitem [{\citenamefont {Zhou}\ and\ \citenamefont {Zettl}(2013)}]{zhou13}%
	\BibitemOpen
	\bibfield  {author} {\bibinfo {author} {\bibfnamefont {Q.}~\bibnamefont
			{Zhou}}\ and\ \bibinfo {author} {\bibfnamefont {A.}~\bibnamefont {Zettl}},\
	}\href {\doibase 10.1063/1.4806974} {\bibfield  {journal} {\bibinfo
			{journal} {Applied Physics Letters}\ }\textbf {\bibinfo {volume} {102}},\
		\bibinfo {pages} {223109} (\bibinfo {year} {2013})}\BibitemShut {NoStop}%
	\bibitem [{\citenamefont {Bunch}\ \emph {et~al.}(2008)\citenamefont {Bunch},
		\citenamefont {Verbridge}, \citenamefont {Alden}, \citenamefont {van~der
			Zande}, \citenamefont {Parpia}, \citenamefont {Craighead},\ and\
		\citenamefont {McEuen}}]{bunch08}%
	\BibitemOpen
	\bibfield  {author} {\bibinfo {author} {\bibfnamefont {J.~S.}\ \bibnamefont
			{Bunch}}, \bibinfo {author} {\bibfnamefont {S.~S.}\ \bibnamefont
			{Verbridge}}, \bibinfo {author} {\bibfnamefont {J.~S.}\ \bibnamefont
			{Alden}}, \bibinfo {author} {\bibfnamefont {A.~M.}\ \bibnamefont {van~der
				Zande}}, \bibinfo {author} {\bibfnamefont {J.~M.}\ \bibnamefont {Parpia}},
		\bibinfo {author} {\bibfnamefont {H.~G.}\ \bibnamefont {Craighead}}, \ and\
		\bibinfo {author} {\bibfnamefont {P.~L.}\ \bibnamefont {McEuen}},\ }\href
	{http://dx.doi.org/10.1021/nl801457b} {\bibfield  {journal} {\bibinfo
			{journal} {Nano Letters}\ }\textbf {\bibinfo {volume} {8}},\ \bibinfo {pages}
		{2458} (\bibinfo {year} {2008})}\BibitemShut {NoStop}%
	\bibitem [{\citenamefont {Davidovikj}\ \emph {et~al.}(2017)\citenamefont
		{Davidovikj}, \citenamefont {Scheepers}, \citenamefont {van~der Zant},\ and\
		\citenamefont {Steeneken}}]{davidovikj17}%
	\BibitemOpen
	\bibfield  {author} {\bibinfo {author} {\bibfnamefont {D.}~\bibnamefont
			{Davidovikj}}, \bibinfo {author} {\bibfnamefont {P.~H.}\ \bibnamefont
			{Scheepers}}, \bibinfo {author} {\bibfnamefont {H.~S.~J.}\ \bibnamefont
			{van~der Zant}}, \ and\ \bibinfo {author} {\bibfnamefont {P.~G.}\
			\bibnamefont {Steeneken}},\ }\href {http://dx.doi.org/10.1021/acsami.7b17487}
	{\bibfield  {journal} {\bibinfo  {journal} {ACS Applied Materials \&
				Interfaces}\ }\textbf {\bibinfo {volume} {9}},\ \bibinfo {pages} {43205}
		(\bibinfo {year} {2017})}\BibitemShut {NoStop}%
	\bibitem [{\citenamefont {Chen}\ \emph {et~al.}(2009)\citenamefont {Chen},
		\citenamefont {Rosenblatt}, \citenamefont {Bolotin}, \citenamefont {Kalb},
		\citenamefont {Kim}, \citenamefont {Kymissis}, \citenamefont {Stormer},
		\citenamefont {Heinz},\ and\ \citenamefont {Hone}}]{chen09}%
	\BibitemOpen
	\bibfield  {author} {\bibinfo {author} {\bibfnamefont {C.}~\bibnamefont
			{Chen}}, \bibinfo {author} {\bibfnamefont {S.}~\bibnamefont {Rosenblatt}},
		\bibinfo {author} {\bibfnamefont {K.~I.}\ \bibnamefont {Bolotin}}, \bibinfo
		{author} {\bibfnamefont {W.}~\bibnamefont {Kalb}}, \bibinfo {author}
		{\bibfnamefont {P.}~\bibnamefont {Kim}}, \bibinfo {author} {\bibfnamefont
			{I.}~\bibnamefont {Kymissis}}, \bibinfo {author} {\bibfnamefont {H.~L.}\
			\bibnamefont {Stormer}}, \bibinfo {author} {\bibfnamefont {T.~F.}\
			\bibnamefont {Heinz}}, \ and\ \bibinfo {author} {\bibfnamefont
			{J.}~\bibnamefont {Hone}},\ }\href {http://dx.doi.org/10.1038/nnano.2009.267}
	{\bibfield  {journal} {\bibinfo  {journal} {Nature Nanotechnology}\ }\textbf
		{\bibinfo {volume} {4}},\ \bibinfo {pages} {861} (\bibinfo {year}
		{2009})}\BibitemShut {NoStop}%
	\bibitem [{\citenamefont {Davidovikj}\ \emph {et~al.}(2018)\citenamefont
		{Davidovikj}, \citenamefont {Bouwmeester}, \citenamefont {van~der Zant},\
		and\ \citenamefont {Steeneken}}]{davidovikj18pumps}%
	\BibitemOpen
	\bibfield  {author} {\bibinfo {author} {\bibfnamefont {D.}~\bibnamefont
			{Davidovikj}}, \bibinfo {author} {\bibfnamefont {D.}~\bibnamefont
			{Bouwmeester}}, \bibinfo {author} {\bibfnamefont {H.~S.~J.}\ \bibnamefont
			{van~der Zant}}, \ and\ \bibinfo {author} {\bibfnamefont {P.~G.}\
			\bibnamefont {Steeneken}},\ }\href {http://dx.doi.org/} {\bibfield  {journal}
		{\bibinfo  {journal} {Proceedings IEEE MEMS}\ } (\bibinfo {year}
		{2018})}\BibitemShut {NoStop}%
\end{thebibliography}
\end{document}